\documentclass[aps,prl,twocolumn,superscriptaddress]{revtex4-2}

\usepackage{color}

\usepackage{mathtools} 
\usepackage{graphicx}
\usepackage{epstopdf}
\usepackage{amsmath}
\usepackage{dcolumn}
\usepackage{bm}
\usepackage{bbold}
\usepackage{amssymb}
\usepackage{mathrsfs}
\usepackage{amsfonts}
\usepackage{braket}
\usepackage{tabularx}
\usepackage{setspace}
\usepackage{graphicx}
\usepackage{verbatim}
\usepackage{caption}
\usepackage{subcaption}
\usepackage{appendix}
\usepackage{esint}
\usepackage[font={small,it}]{caption}
\usepackage{float}
\usepackage{array}

\begin{document}

\title{Comparing Instrument Spectral Sensitivity of Dissimilar Electromagnetic Haloscopes to Axion Dark Matter and High Frequency Gravitational Waves}

\author{Michael E. Tobar}
\email{michael.tobar@uwa.edu.au}
\affiliation{Quantum Technologies and Dark Matter Labs, Department of Physics,  University of Western Australia, 35 Stirling Hwy, 6009 Crawley, Western Australia.}
\author{Catriona A. Thomson}
\affiliation{Quantum Technologies and Dark Matter Labs, Department of Physics,  University of Western Australia, 35 Stirling Hwy, 6009 Crawley, Western Australia.}\author{William M. Campbell}
\affiliation{Quantum Technologies and Dark Matter Labs, Department of Physics,  University of Western Australia, 35 Stirling Hwy, 6009 Crawley, Western Australia.}
\author{Aaron Quiskamp}
\affiliation{Quantum Technologies and Dark Matter Labs, Department of Physics,  University of Western Australia, 35 Stirling Hwy, 6009 Crawley, Western Australia.}
\author{Jeremy F. Bourhill}
\affiliation{Quantum Technologies and Dark Matter Labs, Department of Physics,  University of Western Australia, 35 Stirling Hwy, 6009 Crawley, Western Australia.}
\author{Benjamin T. McAllister}
\affiliation{Quantum Technologies and Dark Matter Labs, Department of Physics,  University of Western Australia, 35 Stirling Hwy, 6009 Crawley, Western Australia.}
\author{Eugene N. Ivanov}
\affiliation{Quantum Technologies and Dark Matter Labs, Department of Physics,  University of Western Australia, 35 Stirling Hwy, 6009 Crawley, Western Australia.}
\author{Maxim Goryachev}
\affiliation{Quantum Technologies and Dark Matter Labs, Department of Physics,  University of Western Australia, 35 Stirling Hwy, 6009 Crawley, Western Australia.}


\begin{abstract}
It is known that haloscopes that search for dark matter axions via the axion-photon anomaly are also sensitive to gravitational radiation through the inverse Gertsenshtein effect. Recently this way of searching for high frequency gravitational waves has gained momentum as it has been shown that the strain sensitivities of such detectors, are of the same order of sensitivity to the axion-photon theta angle. Thus, after calculating the sensitivity of a haloscope to an axion signal, we also have calculated the order of magnitude sensitivity to a gravitational wave signal of the same spectral and temporal form. However, it is unlikely that a gravitational wave and an axion signal will be of the same form, since physically the way the signals are generated are completely different. For gravitational wave detection, the spectral strain sensitivity is in units strain per square root Hz, is the natural way to compare the sensitivity of gravitational wave detectors due to its independence on the gravitational wave signal. In this work, we introduce a systematic way to calculate the spectral sensitivity of an axion haloscope, so instrument comparison may be achieved independent of signal assumptions and only depends on the axion to signal transduction sensitivity and noise in the instrument. Thus, the calculation of the spectral sensitivity not only allows the comparison of dissimilar axion detectors independent of signal, but also allows us to compare the order of magnitude gravitational wave sensitivity in terms of spectral strain sensitivity, allowing comparisons to standard gravitational wave detectors based on optical interferometers and resonant-mass technology.
\end{abstract}

\maketitle

\section{Introduction}

Gravitational wave (GW) astronomy became a reality in 2015~\cite{GWD2016,GWDb2016} and with the continued success and discovery of astronomical sources with the large optical interferometer detectors~\cite{Davis_2021}. Operating detectors are only capable of probing in a relatively narrow band of frequency (100Hz-1kHz), unlike the electromagnetic spectrum, where observations may be conducted over a vast frequency range (from low RF to X-ray). Furthermore, recently High Frequency Gravitational Waves (HFGW) in the MHz-GHz frequency range have been considered as a probe for new physics~\cite{Cruise_2012,Ejlli:2019aa,Ito:2020aa,Aggarwal:2021wy}. This work relates to the comparison of different detector technology capable of exploring the existence of astrophysical sources of HFGWs. Currently there are a variety of solutions proposed and in some cases prototype experiments have already been undertaken based on resonant-mass~\cite{Goryachev2014GW,Goryachev2021GW}, optical interferometers~\cite{Chou2017,Vermeulen_2021}, and the inverse Gertsenshtein effect~\cite{Gertsenshtein61,Boccaletti:1970wm,Fuzfa2016}. In particular, recent work based on the inverse Gertsenshtein effect shows that dark matter axion haloscopes, which search for axions via the two-photon chiral anomaly~\cite{Sikivie83haloscope}, have a corresponding similar order of sensitivity to high frequency gravitational waves~\cite{Clesse2021,berlin2021detecting,domcke2022novel,sokolov2022gravitational}, so that the dimensionless strain sensitivity, $h_g$, to gravitational waves, is of the same order of sensitivity to axion dark matter in terms of the dimensionless axion-photon theta angle, $\theta_a$, so that $h_g\sim\theta_a$~\cite{berlin2021detecting,sokolov2022gravitational}. Note, that $\theta_a$ is related to the product of the axion two-photon coupling, $g_{a\gamma\gamma}$ and the axion scalar field, $a$, by $\theta_a = g_{a\gamma\gamma}a$. Here $g_{a\gamma\gamma}^2$ is in units of kg per second (or $eV^{-2}$) and $a^2$ is in units second per kg (or $eV^{2}$).

The axions is a hypothetical particle postulated to solve the strong charge-parity problem in quantum chromodynamics, predicted to couple very weakly to known particles and to be produced in the early universe, and thus the scientific case that dark matter may include axions or axionlike particles of varied mass and photon coupling has recently gained momentum~\cite{PQ1977,Wilczek1978,Weinberg1978,wisps,K79,Kim2010,Zhitnitsky:1980tq,DFS81,SVZ80,Dine1983,Preskill1983,Sikivie1983,Sikivie1983b,Svrcek_2006,Arvanitaki10,Higaki_2013,Baumann16,Co2020,Co2020b,Co2021,Oikonomou21,Sikivie2021,Sokolov:2021uv,DILUZIO20201,Rodd2021}. Currently there are many varied ideas and designs for detectors worldwide, which implement the principle of dark matter detection through the electromagnetic anomaly~\cite{Payez_2015,McAllisterFormFactor,Gupta2016,McAllister:2016fux,XSWisp,ABRACADABRA,ABRA21,McAllister2017,MadMax17,Millar_2017,Ioannisian_2017,Majorovits_2020,PhysRevD.96.123008,JEONG2018412,IRASTORZA201889,Oue19,Nagano2019,Goryachev2019,Thomson:2021wk,PhysRevD.96.061102,Henning2019,Liu2019,Marsh2019,Sch_tte_Engel_2021,Lawson2019,Anastassopoulos2017,ZhongHS,Kultask20,TOBAR2020,Gelmini2020,berlin2020axion,Lasenby2020b,Gramolin:2021wm,Abeln:2021us,Cat21,universe7070236,10.1093/ptep/ptab051,Devlin2021,Kwon2021,Backes:2021wd,IWAZAKI2021115298,Chigusa:2021vh,PhysRevD.103.096001,Alvarez-Melcon:2021aa,Alesini22}. This means the comparison of axion detectors have become inexact, as axion haloscopes are usually compared by the limit they put on  $g_{a\gamma\gamma}$, which may include different experimental observables, assume different values of axion dark matter density and coherence, and include varying integration times, leading to exclusion plots and comparison between experiments, which may be misleading. Thus, we introduce a systematic way to calculate the axion theta angle spectral sensitivity per $\sqrt{Hz}$ and compare the performance of a varying range of operational detectors as precision instruments without considering the nature of the axion dark matter signal. 

\section{Spectral Sensitivity of a Photonic Axion Haloscope}

In this section, we introduce the spectral density of photon-axion theta angle noise that assumes nothing about the signal or the way we detect it, including the value of the dark matter density~\cite{TOBAR2020}. This technique only considers the instrument signal transduction along with the noise in the detector itself. To do this we characterise the noise within the detector referred to the mean square of the axion-photon theta angle noise as a spectral density, $S_{\theta_N}$ in units $1/Hz$, related by $\langle\theta_N^2\rangle=\int_{f_1}^{f_2}S_{\theta_N}df$. In general electromagnetic axion haloscopes of varying type measure distinctly different observables. Some examples include, power, frequency, magnetic flux, current, voltage, etcetera, and thus for a generic detector we define a generic experimental observable as, $\mathcal{O}$, and axion transduction as, $\mathcal{K}_{\mathcal{O}}$, so that
\begin{equation}
\left\langle\mathcal{O}\right\rangle=\mathcal{K}_{\mathcal{O}}\left\langle \theta_{a}\right\rangle,
\label{trans}
\end{equation}
where the unit of transduction to the axion theta angle are in the same units as the experimental observable. To improve detector sensitivity, the axion dark matter experimentalist aims to maximise the axion signal transduction. 

Associated with the detector transduction is the random noise in the instrument, which limits the precision of the measurement and is usually limited by Nyquist fluctuations of some sort. We define the spectral density of these observable fluctuations as $S_{\mathcal{O}N}$ in the units of the observable squared per Hz. To improve detector sensitivity, the axion dark matter experimentalists tries to minimise this value. Thus, from the value of instrument transduction and noise, we may define the instruments spectral density of axion-photon theta angle noise with a per root Hz sensitivity given by,
\begin{equation}
\sqrt{S_{\theta_{N}}}=\frac{\sqrt{S_{\mathcal{O}_N}}}{|\mathcal{K}_{\mathcal{O}}|}.
\label{spec}
\end{equation}

Here, $S_{\theta_{N}}$ is the spectral density of theta angle fluctuations squared per Hertz. The value of $S_{\theta_{N}}$ is independent of the axion signal, and only dependent on detector parameters such as the magnitude of the axion to signal transduction and the spectral density of noise in the instrument. Thus, this is a good way to characterise and compare the sensitivity of the instrument independent of the signal, in a similar way that gravitational wave detector are characterised and compared with spectral strain sensitivity per $\sqrt{\text{Hz}}$. In the following sections we calculate and compare the spectral sensitivity for a range of axion detectors.

\subsection*{Consideration of the Signal}

When searching for an axion or gravitational wave signal with an axion haloscope, it helps if the wave form of the signal is known. A standard result in signal detection theory~\cite{WZ1962} states that the signal to noise ratio is optimised by a filter which has a transfer function proportional to the complex conjugate of the signal Fourier transform divided by the total noise spectral density. Thus, if we search for a known signal with a matched filter template, the optimum signal to noise ratio is given by,
\begin{equation}
SNR=\frac{1}{2\pi}\int_{-\infty}^{\infty} \frac{\Theta_a(j\omega)^{2}}{S_{\theta_N}(\omega)} d\omega =4 \int_{0}^{\infty} \frac{\Theta_a(f)^{2}}{S_{\theta_N}^{+}(f)} df.
\end{equation}

Here, $\Theta_a(j\omega)$ is the Fourier transform of the axion signal in theta angle per Hz, and $S_{\theta_N}^{+}$ is the single sided spectral density of instrument theta angle noise as given in Equation~(\ref{spec}), which is related to the double sided spectral density by $S_{\theta_N}^{+}(\omega) = 2S_{\theta_N}(\omega)$. This method is applicable when searching for gravitational waves signals of known form, and could also be implemented when scanning for known dark matter transient signals. To look for such signals a method on how to apply optimum signal templates needs to be developed while scanning over the instrument bandwidth. Such templates could be applied by post processing data that already exists from axion haloscopes, which have already put limits on axion dark matter virialized within the galactic halo.

Non-virialized dark matter is considered coherent and of the form, $\theta_a(t)=\theta_a\cos{\omega_a t}$, where $\theta_a = g_{a\gamma\gamma}a_0$ is the theta angle amplitude and $\omega_a=\frac{m_{a}c^2}{\hbar}$ is the effective axion scalar field angular frequency, which is equivalent to the axion mass, $m_a$. The rms amplitude of the signal is easily calculated in the time domain to be, 
\begin{equation}
\left\langle \theta^{2}_a\right\rangle=\frac{1}{T}\int_{0}^{T} \theta_a^2\cos^2{\omega_a t}~~dt = \frac{\theta_a^2}{2};~~T=\frac{2\pi}{\omega_a}.
\label{rmsCo}
\end{equation}

\begin{figure}[H]
\centering
\includegraphics[width=0.4\columnwidth]{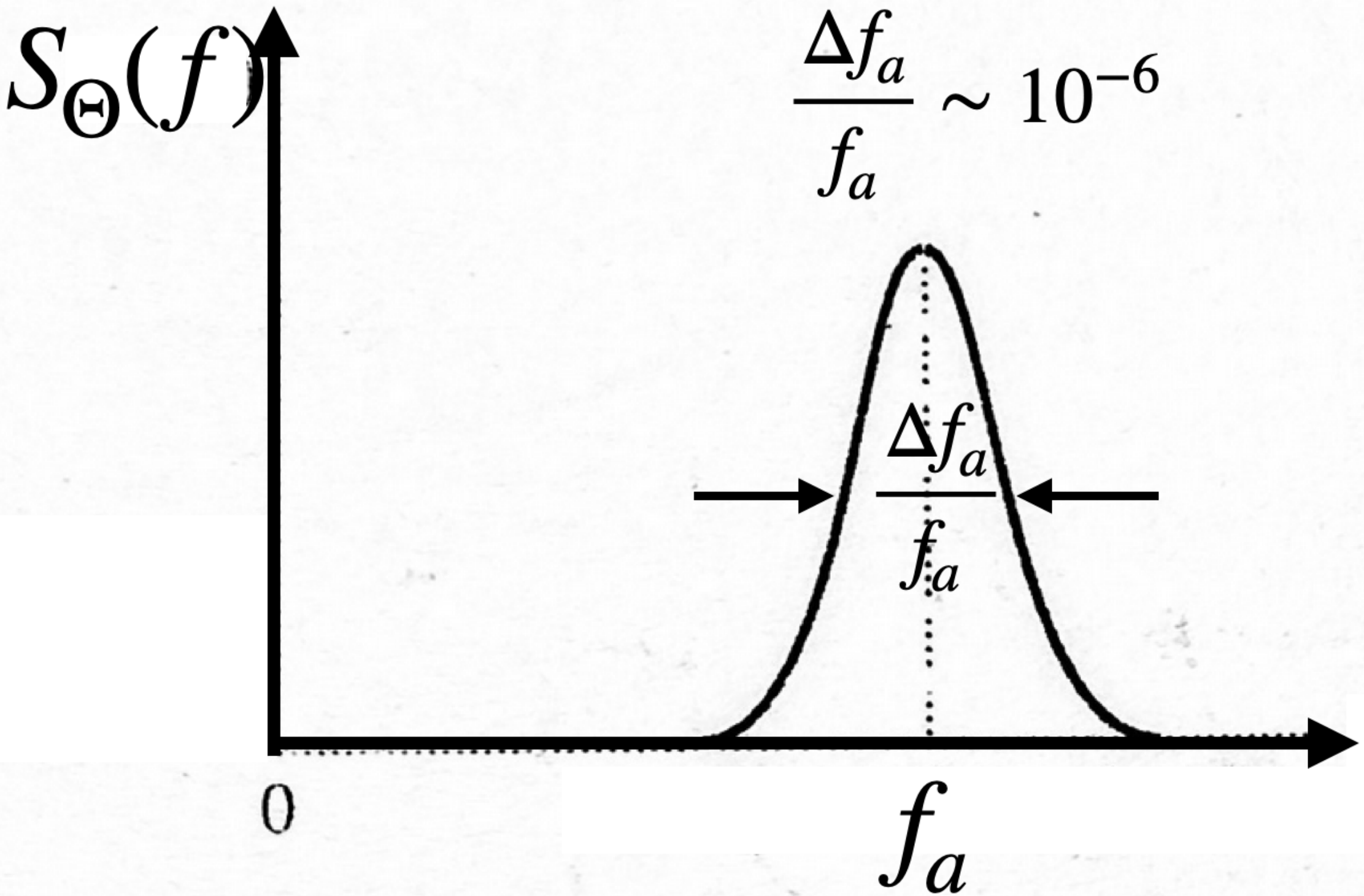}
\caption{Approximation of a virialized dark matter axion as a narrow band noise source as a a Lorentzian distribution and effective Q-factor of $Q_a\sim\frac{f_a}{\Delta f_a}$. In actual fact it takes on a Maxwell–Boltzmann distribution.}
\label{virAx}
\end{figure}

However, when we search for galactic halo dark matter axions, we expect a specific signal in the form of narrow band noise due to the predicted virialization, which is of the form shown in Figure~\ref{virAx}, with a narrowband spectral density, $S_{\Theta}(f)$, in units theta angle squared per Hz. In this case the root mean square of the axion theta angle is given by~\cite{Kim_2020},
\begin{equation}
\left\langle \theta^{2}_a\right\rangle=\int_{-\infty}^{+\infty} \frac{d \omega}{2 \pi}\left|S_\Theta\left(\omega, \omega_{a}\right)\right|^{2}.
\label{rms}
\end{equation}

Note, if we substitute in the Fourier transform of the coherent cosine wave we attain the same value as Equation~(\ref{rmsCo}). However, it has been shown in some circumstance Equation~(\ref{rms}) needs to be used. For example, if the instrument detecting the axion has a narrower bandwidth than the virialized axion itself~\cite{Kim_2020}. Because of the high coherence of the axion, it is usual that the spectral density of instrument noise will be broad compared to the axion. In this case we may use the Dicke radiometer equation \cite{Dicke46}. If the axion coherence time is given by, $\tau_a=\frac{Q_a}{f_a}=\frac{1}{\Delta f_a}$, then for measurement times, $t>\tau_a$, we obtain the following SNR,
\begin{equation}
\sqrt{SNR}=\frac{\mathcal{K}\left\langle \theta_a\right\rangle t^{\frac{1}{4}}\tau_a^{\frac{1}{4}}}{\sqrt{S_{\mathcal{O}_N}}}=\frac{\left\langle \theta_a\right\rangle t^{\frac{1}{4}}\tau_a^{\frac{1}{4}}}{\sqrt{S_{\theta_N}}}.
\end{equation}

For short measurement times, $t<\tau_a$, we may substitute $t^{\frac{1}{4}}\tau_a^{\frac{1}{4}}\rightarrow t^{\frac{1}{2}}$. The strength of the signal in terms of $\left\langle \theta_a\right\rangle$, naturally depends on dark matter density, $\rho_a$, and assuming that the axion is more narrow band than the detector, this is given by,
\begin{equation}
\left\langle \theta_{a}\right\rangle=\sqrt{\left\langle \theta^{2}_a\right\rangle}=g_{a\gamma\gamma}\frac{\sqrt{\rho_a c^{3}}}{\omega_{a}}.
\label{DMdens}
\end{equation}

Given that $\tau_a=\frac{10^6}{f_a}$ we can write the instrument signal to noise ratio to virialized dark matter axions as,
\begin{equation}
\sqrt{SNR}= g_{a\gamma\gamma}  \frac{\sqrt{\rho_a c^{3}}~t^{\frac{1}{4}}\left(\frac{10^6}{f_a}\right)^{\frac{1}{4}}}{2\pi f_a\sqrt{S_{\theta_N}}}.
\label{SNRgen}
\end{equation}

Thus, we have succeeded in writing the SNR of a axion haloscope in terms of the axion-photon theta angle spectral density of noise, $\sqrt{S_{\theta_N}}$. This is general for any axion haloscope which utilises the axion-photon anomaly. If the axion detector is more narrow band than the signal it is detecting, this means there is a modification of the value of $\left\langle \theta_{a}\right\rangle$~\cite{Kim_2020}, not $\sqrt{S_{\theta_N}}$, as the latter is independent of the signal.

In principle there can be more than one component of noise, for example the Nyquist noise generated from the resonator and amplifier. In this case the noise sources should be added in quadrature, assuming $i$ noise sources this can be written as,
\begin{equation}
\sqrt{S_{\theta_N}(f)}=\sqrt{\sum_{i} S^2_{\theta_{Ni}}(f)}=\frac{1}{|\mathcal{K}_\mathcal{O}|}\sqrt{\sum_{i} S^2_{\mathcal{O}_{Ni}}(f)},
\label{multiple}
\end{equation}

\begin{figure}[H]
\centering
\includegraphics[width=0.7\columnwidth]{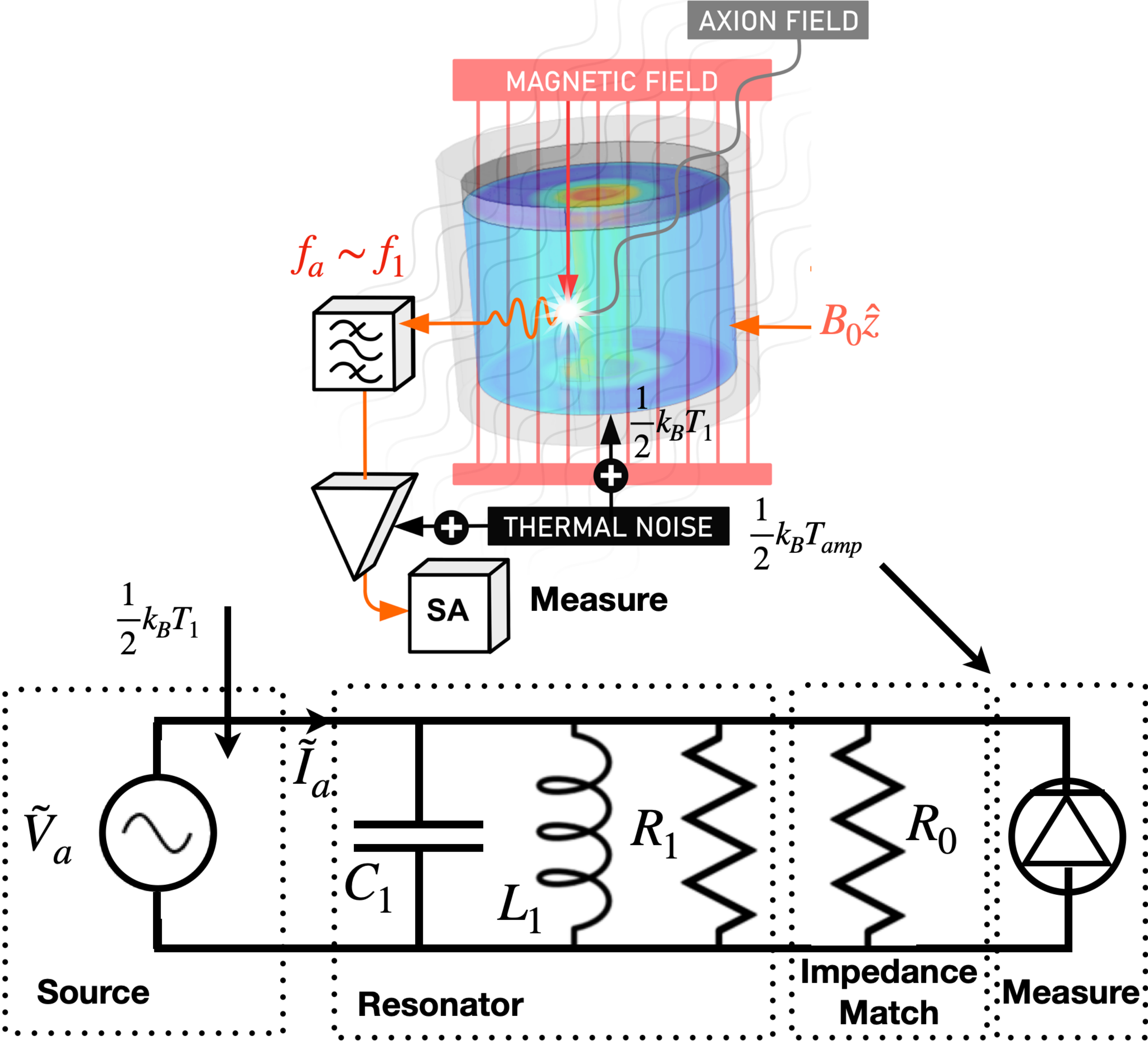}
\caption{Schematic of a resonant cavity haloscope (above) and equivalent circuit (below). Resonant conversion of the axion mixing with the DC magnetic field, $B_0\hat{z}$, creates a photon of frequency $f_a$ near the cavity resonance frequency, $f_1$. The observable signal, $\sqrt{P_a}=\sqrt{\text{Re}(\tilde{V}_a\tilde{I}_a)}$ Equation~(\ref{RHCObs}) competes with the thermal noise sources given in Equation~(\ref{PNrch}).}
\label{RCH}
\end{figure}

\section{Comparison of Resonant Cavity Haloscopes}

In this section, we calculate and compare the spectral sensitivity of some resonant cavity axion haloscopes. A cavity resonator can be either modelled as a series LCR circuit driven by a voltage source or a parallel LCR circuit driven by a current source~\cite{Montgomery87,Dicke1987,Tobar91}. The parallel equivalent circuit is shown in  Figure~\ref{RCH}, where the coupling, $\beta_1$ is defined as the impedance ratio $\beta_1=\frac{R_1}{R_0}$, the resonant frequency as $\omega_1=\sqrt{\frac{1}{L_1C_1}}$ and loaded Q-factor $Q_1=(R_1\parallel R_0)\sqrt{\frac{C_1}{L_1}}$. The masurement observable is the square root power $\sqrt{P_{a}}$ created in the resonant cavity haloscope due to the axion, so Equation~(\ref{trans}) may be written as~\cite{Tobar2022},
\begin{equation}
\sqrt{P_{a}}=\mathcal{K}_{RCH}\left\langle \theta_a\right\rangle.
\label{RHCObs}
\end{equation}

In terms of the microwave parameters, the effective transduction, $\mathcal{K}_{RCH}$, which includes the resonant Lorentzian line-shape, may be written as~\cite{Sikivie83haloscope,Sikivie2021,Kim_2020,Tobar2022}, 
\begin{equation}
\mathcal{K}_{RCH}=\frac{\sqrt{\beta_1\omega_{a} Q_{L1} \epsilon_{0} VC_{01}}cB_{0}}{\sqrt{1+\beta_1}\sqrt{1+4Q^2_1(\frac{\omega_a-\omega_1}{\omega_a})^2}}.
\end{equation}

Here, $V$ is the cavity volume, $C_{01}$ the cavity-axion form factor, and $B_0$ the DC magnetic field strength applied to the cavity. Furthermore, we define the noise spectral density of power fluctuations in the haloscope. For the resonant cavity there are two major terms (see Figure~\ref{RCH}). First, the noise power dominated by thermal noise in the resonator mode of effective temperature, $T_1$, the second is the noise temperature of the first amplifier, $T_{amp}$, which occurs after the resonator, given by
\begin{equation}
\begin{aligned}
P_N\sim\frac{4 \beta_1 }{(\beta_1 +1)^2 \left(1+4 Q_{L1}^2\big(\frac{\omega_a-\omega_1}{\omega_a}\big)^2\right)} \frac{k_BT_{1}}{2}+\frac{k_BT_{amp}}{2}.
\end{aligned}
\label{PNrch}
\end{equation}

Thus, we have two noise sources ($i=2$) that can be added in quadrature as presented in Equation~(\ref{multiple}), so that the spectral theta angle sensitivity becomes,
\begin{equation}
\begin{aligned}
&\sqrt{S_{\theta_{RCH}}}=\frac{\sqrt{k_B\mu_0}}{B_{0}\sqrt{Q_{L1}VC_{01}\omega_a}}\times \\
&\sqrt{\frac{2T_{1}}{(\beta_1+1)}+\frac{T_{amp}(\beta_1+1)}{2\beta_1}\left(1+4 Q_{L1}^2\big(\frac{\omega_a-\omega_1}{\omega_a}\big)^2\right)}.
\end{aligned}
\label{SpecN}
\end{equation}

Note, in general other noise terms exist due to reflections and back action from the amplifier, and interference from thermal noise from any circulators between the cavity and amplifier. However, these noise sources effectively modify either the narrow or broad band noise terms, and thus the effective noise temperature, which is determined experimentally. The first term under the square root sign is the noise temperature due to the Nyquist noise in the resonator, and is a broadband, as the axion signal and this noise term see the same transfer function. The second term is the Nyquist noise added in series, which is dominated by the amplifier and restricts the sensitivity to be around the resonance frequency. The lower this term the bigger the detector bandwidth without tuning, given in (\ref{BW}).

\begin{figure}[H]
\centering
\includegraphics[width=0.8\columnwidth]{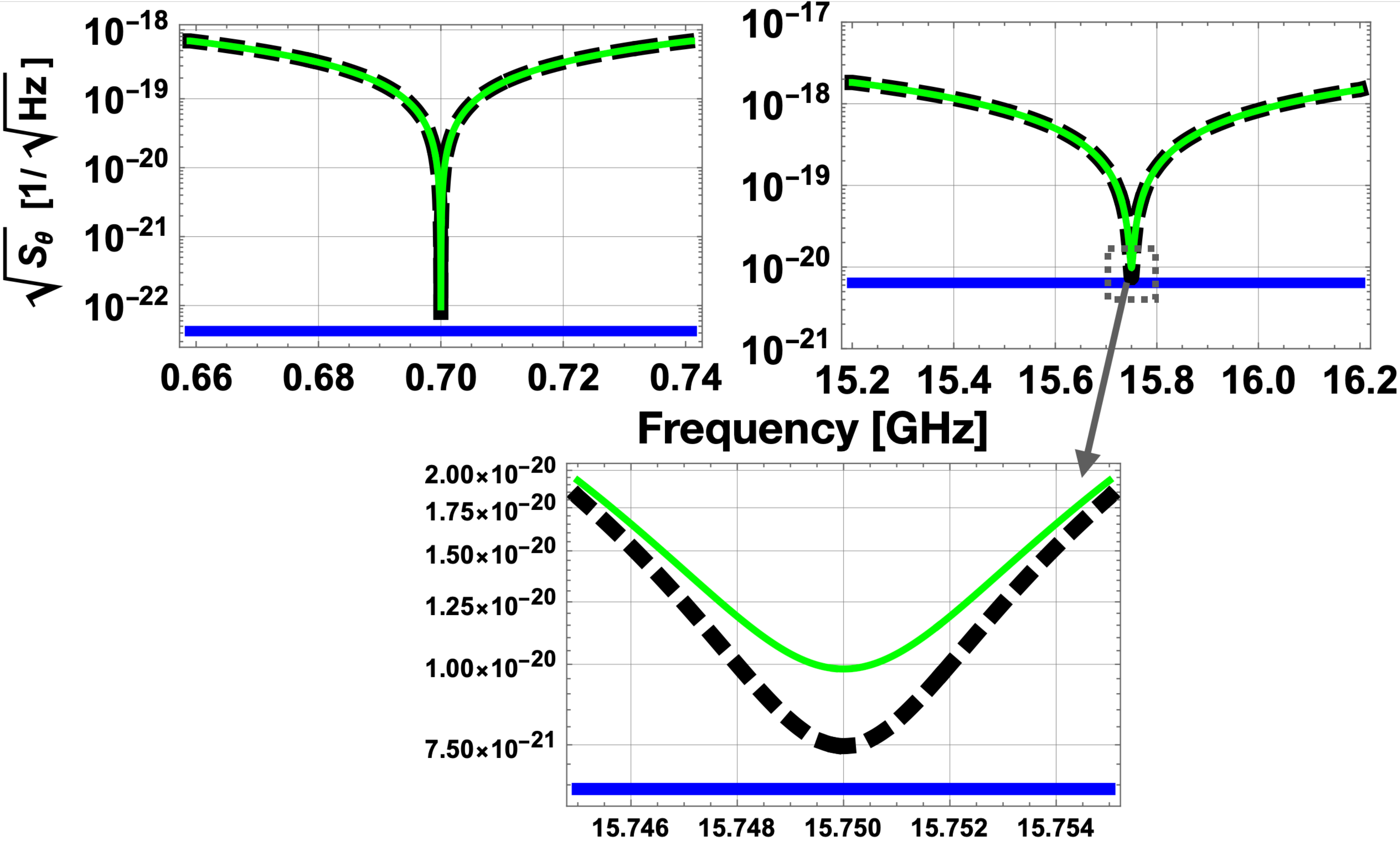}
\caption{Estimated spectral sensitivity for ADMX (\textbf{upper left}) and ORGAN (\textbf{upper right}) from their latest experimental runs~\cite{ADMXCollabPRL2021,Quiskamp2022}. The blue lines shows the cavity thermal noise from the first term in Equation~(\ref{SpecN}) and is broadband. The black dashed curves give the amplifier noise, the second term in Equation~(\ref{SpecN}) and is narrowband. The green curve gives the total noise from the total value of Equation~(\ref{SpecN}), note the amplifier noise dominates off-resonance. Only close to resonance is the cavity thermal noise significant, at this frequency the sum of the two effective noise temperatures adds to limit the sensitivity. This is highlighted for ORGAN (\textbf{lower middle}), which shows a close up near resonance.}
\label{RCHSS}
\end{figure}

In the most recent run of ADMX~\cite{ADMXCollabPRL2021}, $Q_{L1}=8\times10^4$, $V=136$ L, $B_{0}=7.5$ Tesla, $C_{01}\sim0.4$, $T_1$~=~150 mK, $T_{amp}$~=~450 mK and $\beta_1>2$, so $\mathcal{K}_{RCH}\sim10^6\sqrt{f_a}$. The calculated spectral sensitivity is plotted in Figure~\ref{RCHSS} close to 700 MHz, showing the thermal noise from the cavity (broadband noise component), while noise from the amplifier, causes the sensitivity to be narrow band. In contrast, in the most recent run of ORGAN~\cite{Quiskamp2022}, $Q_{L1}=3.5\times10^3$, $V=60$ millilitres, $B_{0}=11.5$ Tesla, $C_{01}\sim0.4$, $T_1$~=~5.2 K, $T_{amp}$~=~5.3 K and $\beta_1>2$,  so $\mathcal{K}_{RCH}=6.5\times10^3\sqrt{f_a}$. The calculated spectral sensitivity is plotted in Figure~\ref{RCHSS} close to 15.75 GHz. From the spectral plots, one can notice the effective bandwidth of the detector while not tuning can be much larger than the cavity bandwidth if the cavity Nyquist noise dominates, and equal to the bandwidth of the resonator if the series amplifier noise dominates. Following the same technique for resonant bars gravitational wave detectors, the effective bandwidth, $\Delta f_{RCH}$, may be calculated from~\cite{Tobar93},
\begin{equation}
\begin{aligned}
\Delta f_{RCH}=
S_{\theta_{RCH}}\left(f_1\right) \int_0^{\infty} \frac{1}{S_{\theta_{RCH}}(f)} df.
\end{aligned}
\label{BW}
\end{equation}

Applying (\ref{BW}) to ADMX and ORGAN, the effective detector bandwidths can be calculated to be 15.87 kHz and 7.1 MHz, respectively, compared to the resonator bandwidths of  8.75 kHz and 4.5 MHz. The detector bandwidth limits the sensitivity when spectral components of the signal lie outside the detector bandwidth. For example, if a transient axion or HFGW pulse strikes the detector with signal strength $\Theta_a(j\omega)$ per Hz, central frequency $f_1$, signal bandwidth of $\Delta f_S$ and pulse width of $\frac{1}{\Delta f_S}$, then the sensitivity of detection if  $\Delta f_S<\Delta f_{RCH}$ with SNR~=~1 and assuming an optimal filter~\cite{Tobar93} is $\langle\theta\rangle=\frac{\sqrt{\Delta f_SS_{\theta_{RCH}}(f_1)}}{2}$. However, if $\Delta f_S>\Delta f_{RCH}$ the sensitivity of detection with SNR~=~1 is $\langle\theta\rangle=\frac{\Delta f_S\sqrt{S_{\theta_{RCH}}(f_1)}}{2\sqrt{\Delta f_{RCH}}}$ , so the sensitivity of detection of the narrow band detector with respect to the broadband detector is degraded by the factor of $\sqrt{\frac{\Delta f_{S}}{\Delta f_{RCH}}}$. If the source is continuous (not transient), then tuning the resonant haloscope to some extent can compensate for its non-broadband nature, but is more complicated and beyond the scope of this work to~discuss.

\section{Comparison of Broadband Reactive Haloscopes}

Broadband reactive haloscopes typically search for axions below 10~MHz or \mbox{$4\times10^{-8}$~eV}. The most common are inductive, which couple through the axion current. Established examples include SHAFT~\cite{Gramolin:2021wm} and ABRACADABRA~\cite{ABRA21} (ABRA for short). Broad band signals occur when the axion velocity is order of the speed of light, in contrast galactic halo axions are considered non-relativistic~\cite{Rodd2021,Fischer2018}. Some possible strategies which could lead to the detection for such broadband signals had been discussed in~\cite{Budker2020}, which can have significant advantage over the narrow band detectors when searching for such signals.

\subsection{ABRACADABRA}
The observable for the ABRA experiment is the axion induced magnetic flux, $\Phi_{a}$ seen by the pick up coil, and is given by
\begin{equation}
\left|\Phi_{a}\right|^{2}= \frac{\omega_{a}^2}{c^{2}} V^{2} \mathcal{G}_{V}^{2} B_{max }^{2}\left\langle \theta_{0}\right\rangle^2.
\end{equation}

Here $V$ is the magnet volume, $\mathcal{G}_{V}$ is the geometric factor that relates the volume to the maximum value of magnetic field in the toroid. Thus, using a sensitive SQUID amplifier in the readout means the transduction from axion theta angle to the magnetic flux observable is given by,
\begin{equation}
\mathcal{K}_{ABRA}=\frac{\omega_{a}}{c} V \mathcal{G}_{V} B_{\max }\frac{M_{in}}{L_{T}}.
\label{abra}
\end{equation}

Here $L_T$ is the total inductance of the pick up coil and the SQUID input impedance, and $M_{in}$ is the mutual inductance between the input inductance and the SQUID, so the axion transduction (\ref{abra}) includes an effective transformer gain of $\frac{M_{in}}{L_{T}}$. The latest experiment has the following design values, $\mathcal{G}_{V}=0.032$, $B_{max}=1\text{T}$, $V=890\times10^{-6}~\text{m}^3$, $L_T=800~\text{nH}$, $M_{in}=2.5~\text{nH}$ so the axion transduction becomes, $\mathcal{K}_{ABRA}=1.9\times 10^{-15}f_a$ Wb.

The SQUID flux noise, $\sqrt{S_{\phi\phi}}$ Wb/$\sqrt{\text{Hz}}$ is commonly represented in terms of the flux quanta, $\Phi_0=\frac{h}{2e}=2.0678\times10^{-15}$ Wb. Fitting the flux noise to the experiment between 70~kHz to 2~MHz~\cite{ABRA21} we obtain $\sqrt{S_{{\phi\phi}_{ABRA}}}\sim\Phi_010^{-6}\sqrt{0.2+\frac{5\times10^{10}}{f_a^2}}$. The spectral theta angle per root Hz can be determined to be,
\begin{equation}
\sqrt{S_{\theta_{ABRA}}}=\frac{cL_T\sqrt{S_{{\phi\phi}_{ABRA}}}}{\omega_{a} V \mathcal{G}_{V} B_{\max }M_{in}}.
\end{equation}

Thus, for the most recent configuration the spectral axion sensitivity becomes $\sqrt{S_{\theta_{ABRA}}}=1.1\times10^{-6}\sqrt{0.2f_a^{-2}+5\times10^{10}f_a^{-4}}$ between 70 kHz to 2 MHz, which is plotted in Figure~\ref{BB}.

\subsection{SHAFT}

The SHAFT experiment also makes use of toroidal magnets but uses permeable cores to enhance the magnetic field at the centre of the magnet and hence enhance the sensitivity. Furthermore, the experiment utilises two detection channels to help veto spurious detections. The transduction of the SHAFT experiment is given by,
\begin{equation}
\mathcal{K}_{SHAFT}=\frac{\omega_{a}}{c} V_{eff}B_{\max }\frac{N_p M_{in}}{L_{T}}.
\label{shaft}
\end{equation}

Here $V_{eff}$ replaces $V \mathcal{G}_{V}$ in (\ref{abra}) and the pick up coil has more than one turn ($N_p$). In this experiment the total inductance includes the inductance of the twisted pair ($L_{tp}$), so $L_T=L_p+L_{tp}+L_{in}$, given that $N_p=6$, $L_p=3.32~\mu\text{H}$, $L_{tp}=0.1~\mu\text{H}$, $L_{in}=1.8~\mu\text{H}$, $M_{in}=8.6~\text{nH}$, $B_{\max}=1.51~\text{T}$, and $V_{eff}=10.3\times10^{-6}$, then $\mathcal{K}_{SHAFT}=3.2\times 10^{-15}f_a$ Wb, 1.7 times bigger than ABRA.

The SQUID flux noise in SHAFT is inferred from the magnetic flux density, fitting to these experimental results between 2kHz to 6MHz, we obtain, $\sqrt{S_{{\phi\phi}_{SHAFT}}}\sim\Phi_010^{-6}\sqrt{0.688+1.76\times10^{30}f_a^{-8}+3.48\times10^{-26}f_a^4}$. The spectral theta angle per root Hz for SHAFT can thus be determined to be,
\begin{equation}
\sqrt{S_{\theta_{SHAFT}}}=\frac{cL_T\sqrt{S_{{\phi\phi}_{SHAFT}}}}{\omega_{a} V_{eff} B_{\max }N_pM_{in}}.
\end{equation}

\begin{figure}[H]
\centering
\includegraphics[width=0.7\columnwidth]{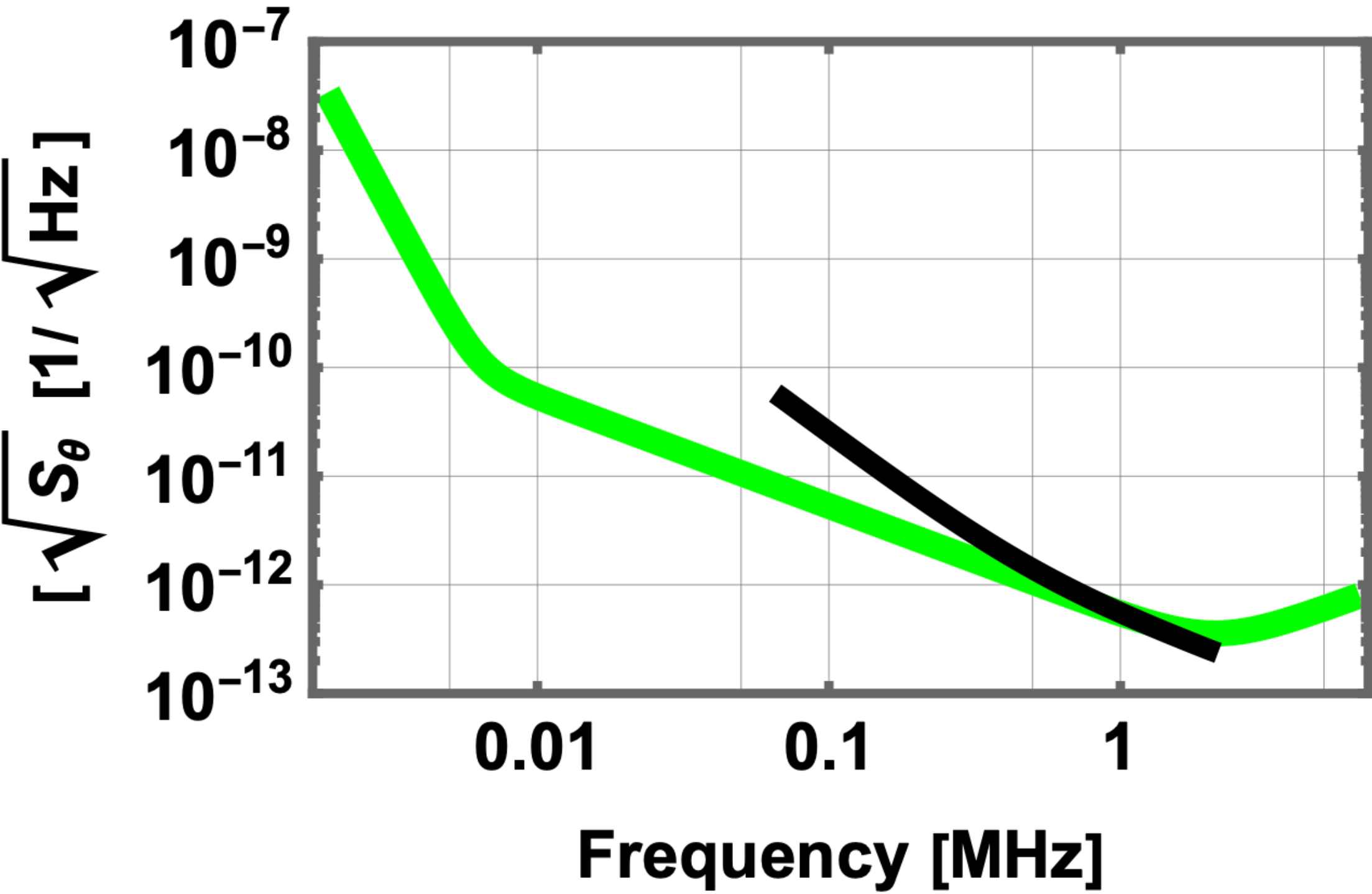}
\caption{Estimated spectral sensitivity for the broadband haloscopes ABRACADABRA (black) and SHAFT (green) based on data in its latest experimental runs~\cite{ABRA21,Gramolin:2021wm}.}
\label{BB}
\end{figure}

Thus, for the most recent configuration the spectral axion sensitivity between 2~kHz to 6~MHz, becomes $\sqrt{S_{\theta_{SHAFT}}}\sim6.4\times10^{-7}\sqrt{0.688f_a^{-2}1.76\times10^{30}f_a^{-10}+3.48\times10^{-26}f_a^2}$, which is plotted in Figure~\ref{BB}. This shows that SHAFT has both a larger transduction from axion to magnetic flux and a lower noise readout compared to ABRA.

\section{Axion Detectors in the 1--500 MHz Band}

ABRA and SHAFT currently only go up to a few MHz in frequency, while the lowest frequency that ADMX put limits on is around 500 MHz. In between only ADMX-SLIC has put significant limits in a narrow band of 42--43 MHz~\cite{ADMXSLIC}. There are various forms of cavity haloscope experiments proposed to search this mass range, and it is possible that DFSZ sensitivity will be achieved in the 100--500 MHz range using novel cavity designs~\cite{LFADMX}. From a few MHz to above a 100 MHz, another option is to implement the dual mode technique~\cite{Cat21}, which uses an oscillating background field rather than a DC background field with prototype experiments already demonstrated. This type of detector has also been proposed as a technique to search for ultra high frequency gravitational waves~\cite{Berlin2022}. Astrophysical sources and detection of gravitational waves above a MHz is relatively unexplored, and the use of axion haloscopes to search for gravitational waves at these frequencies is thought to be a unique opportunity to probe the physics of the very early Universe, as well as beyond standard model physics~\cite{Aggarwal:2021wy}. 

\subsection{Resonant Upconversion Haloscopes}

The upconversion technique usually uses two modes in the same resonant cavity, a background mode, which we label with subscript $0$, and a readout mode, which we label with subscript $1$, with a necessary non zero overlap between the modes over the cavity volume ($V_{cav}$), defined by, $\xi_{10}=\frac{1}{V_{cav}}\int\mathbf{e}_1\cdot\mathbf{b}_0d\tau$, where $\mathbf{e}_1$ is the electric field unit vector of the readout mode and $\mathbf{b}_0$ is the magnetic field unit vector of the background mode~\cite{Goryachev2019,Thomson:2021wk}. This technique allows the upconversion of the axion to the carrier frequency of the readout mode and allows the use of high-Q cavities instead of large magnetic fields, and allows search for axions in the frequency range $\omega_a<<\omega_1$. This technique was first proposed in~\cite{Goryachev2019,Thomson:2021wk}, and showed that a putative dark matter axion back ground would perturb the frequency (or phase) and amplitude (or power) of the readout mode. The former we call the ``frequency technique'' and the later the ``power technique''. 

The first prototype experiment of the ``frequency technique'' has already been performed~\cite{Cat21}, which looked for phase or frequency variations impinged on the read out oscillator. Other variations of the ``power technique'' have recently been proposed~\cite{berlin2020axion,Lasenby2020b,ABerlin2021,Lasenby2020} and performed~\cite{UpconvCat22}, which excite only the background mode, then look for power generated at the readout mode frequency. In addition, recently a new way to implement the upconversion power technique has been proposed, where a mode with non-zero helicity acts as its own background mode, but is only sensitive to amplitude modulations due to ultra-light axions within the resonator bandwidth~\cite{Anyon22}, which is in the regime where the dual-mode upconversion haloscope becomes hard to implement. In this section, we calculate the spectral sensitivity of these techniques.

\subsubsection{Dual-Mode Power Observable}

For the power upconversion haloscope the observable is the square root power $\sqrt{P_{1}}$, of the readout mode so the transduction equation given by (\ref{trans}) is of the form,
\begin{equation}
\sqrt{P_{1}}=\mathcal{K}_{P_{uc}}\left\langle \theta_a\right\rangle,
\end{equation}
with the effective transduction thus given by~\cite{UpconvCat22},
\begin{equation}
\begin{aligned}
&\mathcal{K}_{P_{uc}}=\frac{\xi_{10}2\sqrt{2}\omega_a\sqrt{\beta_{0}Q_{L0}\beta_1Q_{L1}P_{0inc}}}{\sqrt{\omega_1\omega_0}\sqrt{1+\beta_1}(\beta_0+1)\sqrt{1+4Q^2_{L1}\big(\frac{\delta\omega_a}{\omega_1}\big)^2}},
\end{aligned}
\label{PowerSens2}
\end{equation}
where we define $\delta\omega_a=\omega_1+\omega_a-\omega_0$, so when $\delta\omega_a=0$ then $\omega_a=\omega_0-\omega_1$ and the axion induced power is upconverted to the frequency, $\omega_1$. Thus, $\delta\omega_a$ defines the detuning of the induced power with respect to the readout mode frequency.

The noise power is similar to the resonant cavity haloscopes and dominated by thermal noise in the readout mode resonator of effective temperature, $T_1$ and the noise temperature of the first amplifier, $T_{amp}$ after the read out mode, and is given by~\cite{parker2013b},
\begin{equation}
\begin{aligned}
P_N\sim\frac{4 \beta_1 }{(\beta_1 +1)^2 \left(1+4 Q_{L1}^2\big(\frac{\delta\omega_a}{\omega_1}\big)^2\right)} \frac{k_BT_{1}}{2}+\frac{k_BT_{amp}}{2}.
\end{aligned}
\label{PN}
\end{equation}

In the case $\beta_1\sim1$ and $\delta\omega_a\sim0$ then $P_N\sim\frac{k_B(T_1+T_{amp})}{2}$. Thus, from (\ref{spec}), the spectral sensitivity is given by,
\begin{equation}
\begin{aligned}
\sqrt{S_{\theta_{Puc}}}=\frac{\sqrt{\omega_0\omega_1}\sqrt{k_B}(1+\beta_0)}{2\xi_{10}\omega_a\sqrt{2}\sqrt{\beta_1Q_{L1}\beta_0Q_{L0}P_{0_{inc}}}}\times \\\sqrt{\frac{2T_{1}\beta_1}{(\beta_1+1)}+\frac{T_{amp}(\beta_1+1)}{2}\left(1+4 Q_{L1}^2\big(\frac{\delta\omega_a}{\omega_a}\big)^2\right)}.
\end{aligned}
\label{specPUC}
\end{equation}

\begin{figure}[H]
\centering
\includegraphics[width=0.7\columnwidth]{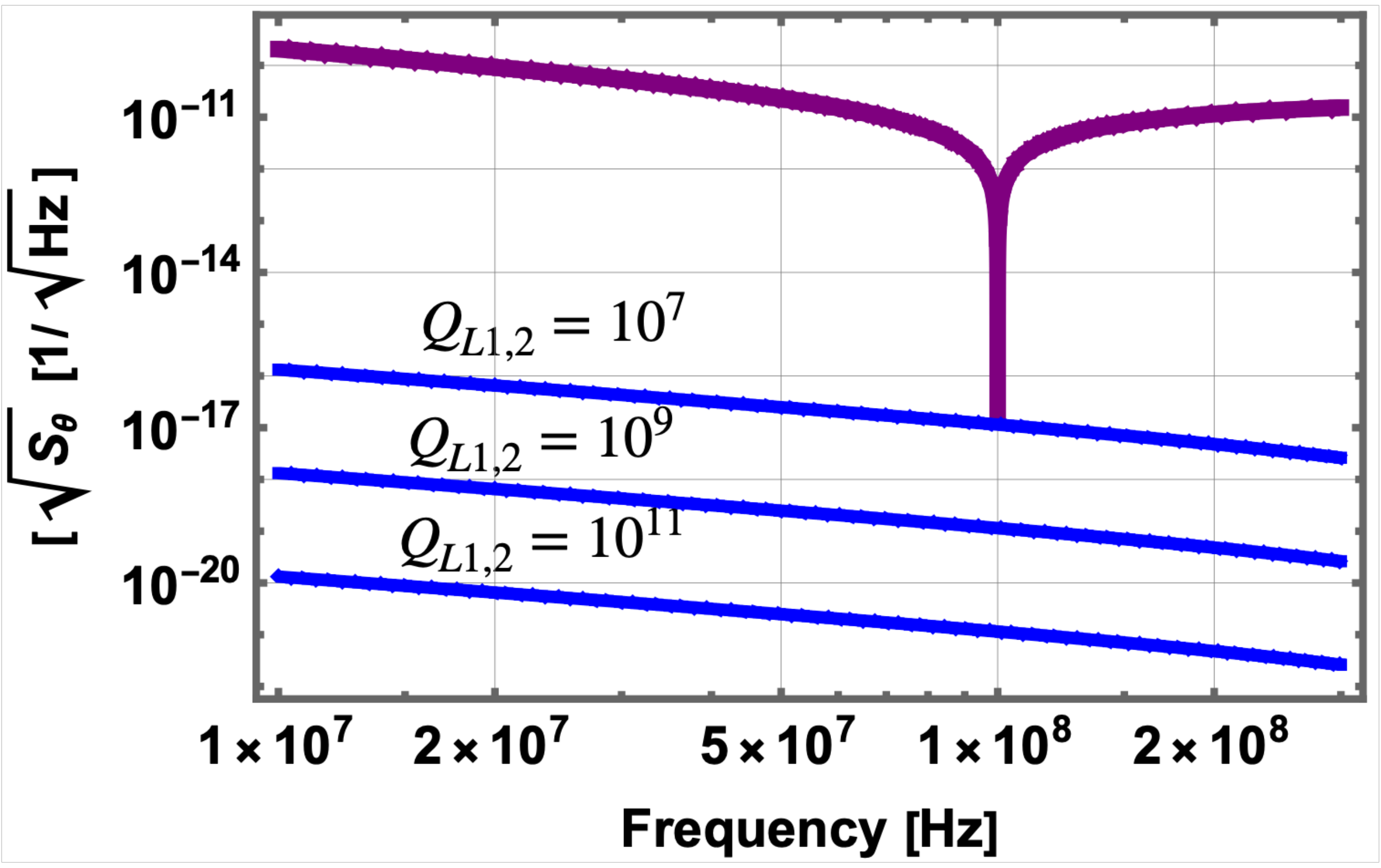}
\caption{Estimated spectral sensitivity for the dual-mode upconversion cavity haloscope. The purple line gives the sensitivity when the readout mode is tuned to 100 MHz away from the background mode, assumed in these simulations to be 1 GHz. The blue curves show the locus of achievable sensitivity as the mode is tuned for a range of cavity mode Q-factors, which have been achieved in prior work using superconducting cavities~\cite{Posen20,Martinello2018}.}
\label{DMode}
\end{figure}

\subsubsection{Dual-Mode Frequency Observable}

For the frequency upconversion haloscope the observable is the rms fractional frequency fluctuations $\langle\frac{\delta\omega_1}{\omega_1}\rangle$, so the transduction equation given by (\ref{trans}) is of the form,
\begin{equation}
\begin{aligned}
\langle\frac{\delta\omega_1}{\omega_1}\rangle
&=\mathcal{K}_{\omega_{uc}}~\left\langle \theta_{a}\right\rangle,
\end{aligned}
\label{PertExp4}
\end{equation}
with the effective transduction thus given by~\cite{UpconvCat22},
\begin{equation}
\mathcal{K}_{\omega_{uc}}=\frac{1}{2}\frac{\omega_a}{\sqrt{\omega_1\omega_0}}\frac{\sqrt{\beta_{0}(\beta_1+1)}\sqrt{Q_{L0}}}{\sqrt{\beta_{1}} \sqrt{Q_{L1}}(\beta_0+1)}\sqrt{\frac{P_{0inc}}{P_{1inc}}}|\xi_{10}|
\label{PertExp5}
\end{equation}

\subsubsection{Single-Mode Anyon Haloscope}

The lowest noise oscillators are frequency stabilized by a phase detection scheme, which in principle is limited by the effective readout system noise temperature $T_{RS}$ of the internal phase detector (includes the amplifier noise temperature), which is close to ambient temperature for a well-designed system~\cite{LONoiseFloor,Ivanov:2006yk,Ivanov:2009pv}, and in such a case the oscillator noise will be,
\begin{equation}
\sqrt{S_{y1}}=\frac{\sqrt{k_{b} T_{R S}}}{\sqrt{2} Q_{L1} \sqrt{P_{1inc}}} \frac{(1+\beta_1)}{2 \beta_1}\sqrt{1+4Q_{L1}^2\left(\frac{\delta\omega_a}{\omega_{1}}\right)^{2}},
\label{sigma2}
\end{equation}
where $P_{1inc}$ is power incident on the input port to the readout mode. 

Thus, from (\ref{spec}), the spectral sensitivity is given by,
\begin{equation}
\begin{aligned}
\sqrt{S_{\theta_{\omega_{uc}}}}=\frac{\sqrt{\omega_0\omega_1}}{\omega_a}\frac{\sqrt{k_B}(1+\beta_0)}{\xi_{10}2\sqrt{2}\sqrt{\beta_1Q_{L1}\beta_0Q_{L0}P_{0_{inc}}}}\times \\
\sqrt{\frac{T_{RS}(\beta_1+1)}{2}\left(1+4 Q_{L1}^2\big(\frac{\delta\omega_a}{\omega_a}\big)^2\right)},
\end{aligned}
\label{specWUC}
\end{equation}
which is of similar form to (\ref{specPUC}) without the broadband term. The broad band term is due to Nyquist fluctuations within the resonator and sees the same cavity transfer function as the signal and may be added in as a second term in a similar way to the power scheme to give the same broadband term. Thus, the frequency and power technique effectively give a similar spectral sensitivity and depend on the noise temperature of the system. 

To calculate the likely sensitivity we implementing Equation (\ref{specPUC}), and assume the use of a superconducting cavity in a dilution refrigerator. This will enable Q-factors of $10^7$--$10^{11}$, depending on the superconducting surface resistance~\cite{Posen20,Martinello2018}, and assuming cavity and amplifier noise temperatures similar to that of the ADMX experiment ($T_1=150~\text{mK}$ and $T_{amp}=450~\text{mK}$), the achievable spectral theta angle sensitivity is plotted in Figure~\ref{DMode}.

\begin{figure}[H]
\centering
\includegraphics[width=0.8\columnwidth]{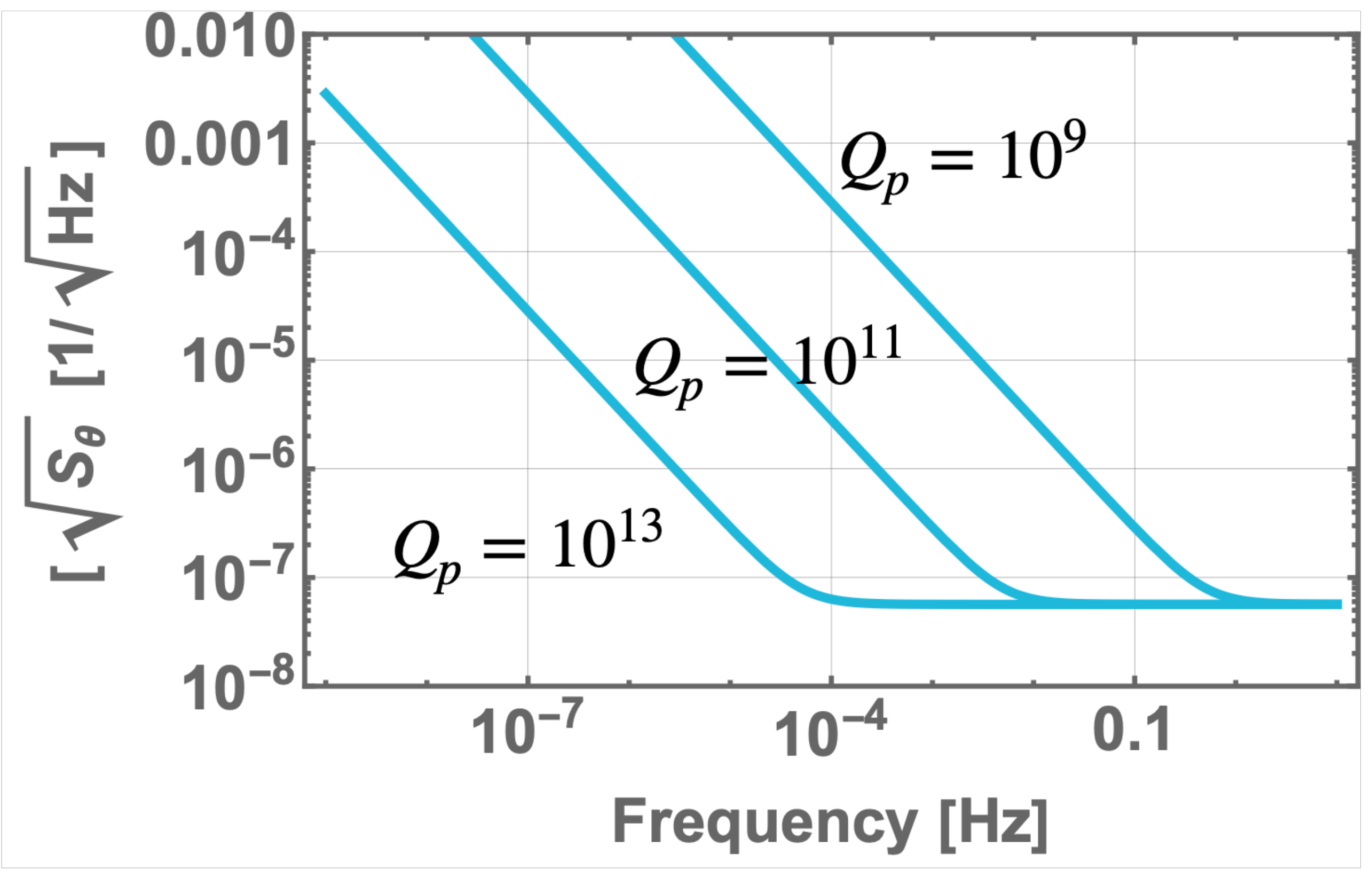}
\caption{Estimated spectral sensitivity for the anyon upconversion cavity haloscope for a range of cavity mode Q-factors. The anyon cavity has a greater Q-factor per surface resistance of material, so based on what has been achieved in prior work up to $10^{13}$ may be possible~\cite{Anyon22}.}
\label{anyon}
\end{figure}

The anyon cavity resonator and its sensitivity to axions has been detailed in~\cite{Anyon22}. This system is of similar to the dual-mode upconversion haloscope except only a single mode with non-zero helicity is required, defined by,
\begin{equation}
\begin{aligned}
\mathscr{H}_p=\frac{2 \operatorname{Im}[\int\mathbf{B}_p(\vec{r})\cdot\mathbf{E}^*_p(\vec{r})~d\tau]}{\sqrt{\int\mathbf{E}_p(\vec{r})\cdot\mathbf{E}_p^*(\vec{r})~d\tau\int\mathbf{B}_p(\vec{r})\cdot\mathbf{B}_p^*(\vec{r})~d\tau}},
\end{aligned}
\label{eq:helicity}
\end{equation}
where $\mathbf{E}_p(\vec{r})$ and $\mathbf{B}_p(\vec{r})$ are the respective mode electric and magnetic field vector phasor amplitudes. The signal to noise ratio for coupling of axions to modes with non zero helicity have been shown to be of the form~\cite{Anyon22},
\begin{equation}
\begin{aligned}
SNR=\frac{g_{a \gamma \gamma}\beta_p|\mathscr{H}_p|}{\sqrt{2}(1+\beta_p)}\frac{Q_p}{\sqrt{1+4Q^2_p(\frac{\omega_a}{\omega_p})^2}}\frac{\left(\frac{10^{6} t}{\omega_{a}}\right)^{\frac{1}{4}} \sqrt{\rho_{a} c^{3}}}{\omega_p\sqrt{S_{am}}},
\end{aligned}
\label{SNRany}
\end{equation}
where $\omega_p$ is the mode frequency, $\beta_p$ the mode coupling, $Q_p$ the loaded Q-factor, and $S_{am}$ represents the spectral density of the pump oscillator amplitude fluctuations driving the helical mode. Comparing with Equation~(\ref{SNRgen}) we identify the transduction with respect to amplitude modulated sidebands to be,
\begin{equation}
\mathcal{K}_{anyon} = \frac{\beta_p}{\sqrt{2}(1+\beta_p)}\frac{Q_p}{\sqrt{1+4Q^2_p(\frac{\omega_a}{\omega_p})^2}}\Big(\frac{\omega_a}{\omega_p}\Big)\lvert\mathcal{H}_p\rvert,
\end{equation}
and thus the spectral strain per root $Hz$ for the anyon haloscope to be,
\begin{equation}
\sqrt{S_{\theta_{anyon}}}= \frac{\sqrt{2}(1+\beta_p)}{\beta_p} \frac{\sqrt{1+4Q^2_p(\frac{\omega_a}{\omega_p})^2}}{Q_p}\Big(\frac{\omega_p}{\omega_a}\Big)\sqrt{S_{am}},
\end{equation}
which is plotted in Figure~\ref{anyon} for achievable experimental values, $\beta_p=1$, $|\mathcal{H}_p|=1$, $\omega_p/2\pi=1$~GHz, and amplitude noise $S_{am}=-160$~dBc/Hz~\cite{rubiolaAM}. Cavity $Q$-factors $Q_p$ are calculated from mode geometry factors, $G_m=Q_pR_s$, where $R_s$ is the surface resistance~\cite{Anyon22,Posen20,Martinello2018}.

\section{Axion Haloscope Sensitivity to Gravitational Waves}

\begin{figure}[H]
\centering
\includegraphics[width=1.0\columnwidth]{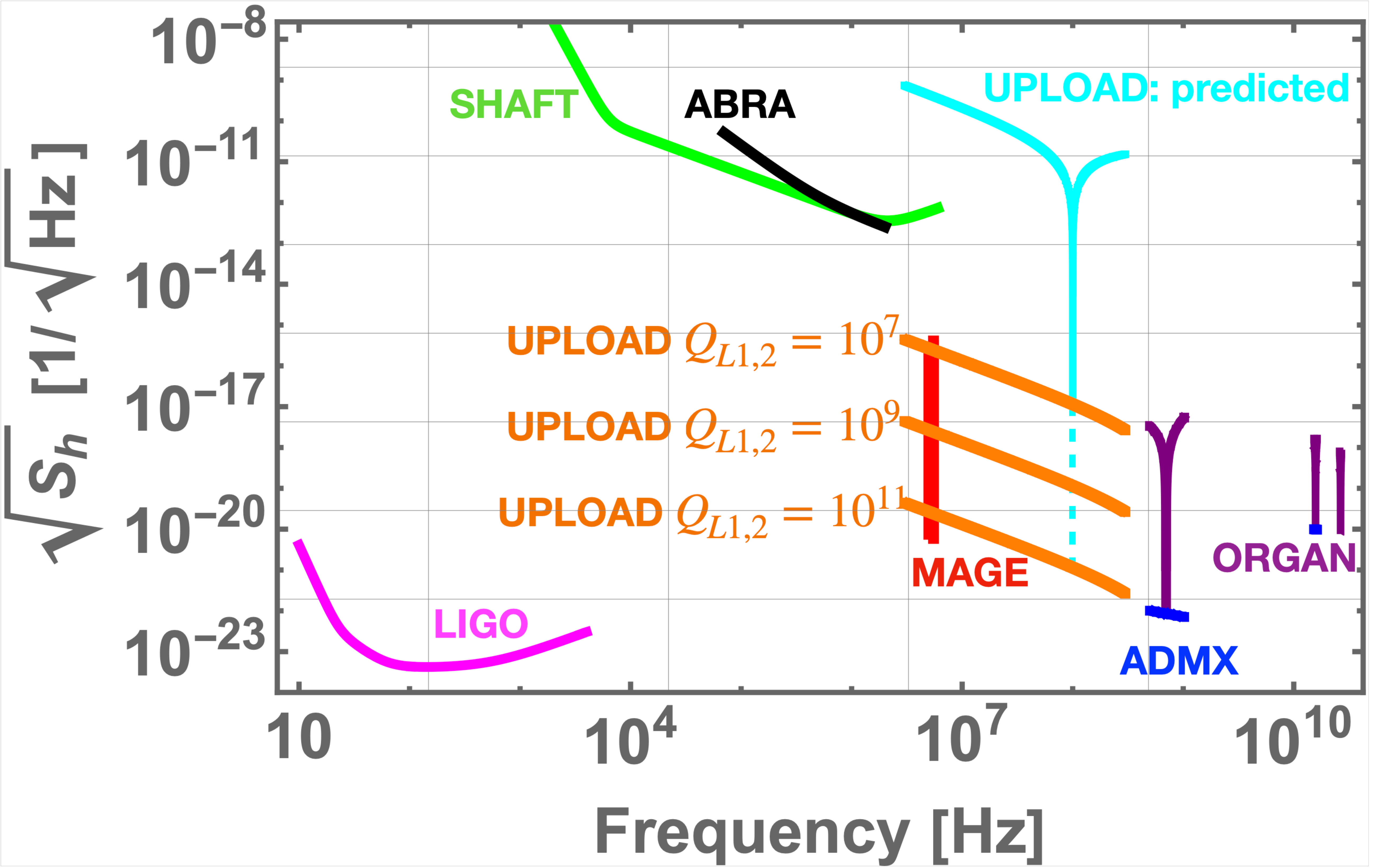}
\caption{Instrument spectral sensitivity, $\sqrt{S_h}$ for the LIGO (10 Hz--4 kHz)~\cite{Davis_2021} and MAGE (4.99 and 5.5 MHz)~\cite{Goryachev2014GW,Goryachev2021GW} gravitational wave detectors, compared with instrument spectral sensitivity, $\sqrt{S_{\theta}}$, for operational axion detectors, SHAFT (2 kHz--6MHz)~\cite{Gramolin:2021wm}, ABRA (70 kHz--2MHz)~\cite{ABRA21}, ADMX (\mbox{0.5--1.2~GHz)}~\cite{ADMXCollabPRL2021} and ORGAN (15.2--16.2 GHz and 26.53~GHz)~\cite{McAllister2017,Quiskamp2022}. We also compare the sensitivity to predictions for $\sqrt{S_{\theta}}$ for the future UPLOAD cryogenic experiment~\cite{Cat21}, which will overlap with frequencies from the MAGE experiment.}
\label{SSGWs}
\end{figure}

Recent research has shown that axion dark matter detectors, which utilise the photon chiral anomaly to convert axions into photons are also sensitive to gravitational waves through the inverse Gertsenshtein effect~\cite{berlin2021detecting,domcke2022novel,Berlin2022}. Assuming, $h_g\sim\theta_a$, we may compare the instrument spectral sensitivity of axion haloscope detectors to current operating detectors such as LIGO~\cite{Davis_2021} and MAGE: Multi-mode Acoustic  Gravitational-wave Experiment~\cite{Goryachev2021GW,MAGE}. This comparison is shown in Figure~\ref{SSGWs}, and if the proportionality between $h_g$ and $\theta_a$ is not exactly unity for an axion halosope, then the axion haloscope sensitivity to gravitational waves may be simply adjusted by this factor.

\section{Discussion}

The gravity wave community usually characterise the instrument sensitivity of their detectors by determining the spectral strain sensitivity of the detectors. This allowed in the past for narrow band resonant bar gravitational antenna to be compared to broadband laser interferometer detectors. We have introduced a similar way of characterising the spectral sensitivity of axion haloscopes, so that broadband reactive haloscopes may be compared to resonant cavity haloscopes without considering the form of the axion dark matter signal. Since it has been shown that axion haloscopes are also sensitive to gravitational waves, we have used this technique to get an idea of the instrument sensitivity to gravitational waves by comparing them together as shown in Figure~\ref{SSGWs}. This has shown that axion detectors have the capability of searching for high frequency gravitational waves, and in the MHz band offer a way to correlate between detectors of different type, such as comparisons of UPLOAD with MAGE or the Fermilab Holometer~\cite{Chou2017}. However, the question remains, are these detectors sensitive enough to detect high frequency gravitational waves? Unlike LIGO, which regularily detects sources in the 100 Hz-1 kHz band, there is a lack of standard sources higher than MHz frequencies. However, exploring these frequencies gives us a chance to explore beyond standard model physics, as outlined in the past~\cite{Cruise_2012}, and a recent white paper on the subject~\cite{Aggarwal:2021wy}. Furthermore, existing and future data from axion detectors can be used to search for HFGWs by just implementing different optimal filters or templates, so the cost of this search only requires additional effort on the data analysis side, and constructing new detectors is not necessary for initial searches at these frequencies. \\

\noindent\textbf{Acknowledgments}

This work was funded by the Australian Research Council Centre of Excellence for Engineered Quantum Systems, CE170100009 and  the Australian Research Council Centre of Excellence for Dark Matter Particle Physics, CE200100008. \\

\providecommand{\noopsort}[1]{}\providecommand{\singleletter}[1]{#1}%

\end{document}